\documentstyle[12pt]{article}
\newlength{\abstractwidth}
\begin{document}
\thispagestyle{empty}
\pagestyle{plain}
\renewcommand{\thefootnote}{\fnsymbol{footnote}}
\renewcommand{\thanks}[1]{\footnote{#1}} 
\newcommand{\starttext}{
\setcounter{footnote}{0}
\renewcommand{\thefootnote}{\arabic{footnote}}}
\newcommand{\be}{\begin{equation}}
\newcommand{\ee}{\end{equation}}


\begin{titlepage}
\bigskip
\hskip 3.7in\vbox{\baselineskip12pt
\hbox{MIT-CTP-2657}\hbox{hep-th/9707135}}
\bigskip\bigskip\bigskip\bigskip

\centerline{\large \bf Interaction of non-parallel D1-branes}

\bigskip\bigskip
\bigskip\bigskip

\centerline{\bf Alec Matusis\thanks{alec\_m@ctp.mit.edu}}
\medskip
\centerline{Center for Theoretical Physics}
\centerline{Massachusetts Institute of Technology}
\centerline{Cambridge, MA\ \ 02139}

\bigskip\bigskip

\begin{abstract}
\baselineskip=16pt
We find the potential per unit length between two non-intersecting D1-branes as a function of their relative angle. 
\end{abstract}
\end{titlepage}
\starttext
\baselineskip=18pt
\setcounter{footnote}{0}
This  calculation is a generalization of the calculation of the amplitude between two parallel D-branes~\cite{pol}. First we quantize a free field theory with the appropriate boundary conditions and then compute one loop annulus amplitude in the operator formalism. This computation resembles the case of the open strings in the background electromagnetic field~\cite{bachas}. However, in our case, there are no boundary terms in the Lagrangian, which described the interaction between the string and the background field. 

Let $X^i(0,\tau)=0, i=2..9$ be the equation describing the position of the first D-string and $X^2(\pi,\tau)-X^1(\pi,\tau)\tan \alpha=0, X^j(\pi,\tau)=Y, j=3..9$ be the equation for the second one. In this notation, $\alpha$ is the relative angle and $Y$ is the minimal separation between D-strings.The boundary conditions then look like \begin{eqnarray}
&{\partial}_{\sigma}X^1(0,\tau) = 0,\nonumber \\
& X^2(0,\tau)=0,\nonumber \\
& {\partial}_{\sigma}X^1(\pi,\tau)+{\partial}_{\sigma}X^2(\pi,\tau)\tan \alpha = 0,\nonumber \\
& X^2(\pi,\tau)-X^1(\pi,\tau)\tan \alpha=0.\end{eqnarray} 
From the boundary conditions and the wave equation we have \begin{eqnarray} 
 &\dot{X}^1_L(\tau +\pi)-\dot{X}^1_R(\tau -\pi)=-\tan \alpha (\dot{X}^2_L(\tau +\pi)-\dot{X}^2_R(\tau -\pi))& \nonumber \\
 &(\dot{X}^1_L(\tau +\pi)+\dot{X}^1_R(\tau -\pi))\tan \alpha =\dot{X}^2_L(\tau +\pi)+\dot{X}^2_R(\tau -\pi)&\nonumber \\
 &X^2_R(\tau)+X^2_L(\tau)=0&\nonumber \\
 &\dot{X}^1_R(\tau)-\dot{X}^1_L(\tau)=0,\label{eq:bbound}\end{eqnarray}
where dot means the derivative with respect to the argument.
These functional equations can be solved for $\dot{X}_{L,R}$ and integrated, yielding the mode expansion:
\begin{eqnarray}&& X^1=\frac{1}{\sqrt2}\sum_{n\in \cal Z}[a_n \psi_n + h.c.]\nonumber\\
 &&X^2=\frac{1}{\sqrt2}\sum_{n\in \cal Z}[-ia_n \phi_n + h.c.],\end{eqnarray}  where the functions $\psi_n $and $\phi_n $ are \begin{eqnarray}
 \psi_n=(n+\frac{\alpha}{\pi})^{-\frac{1}{2}}e^{i\tau(n+\frac{\alpha}{\pi})}\cos \left[\sigma (n+\frac{\alpha}{\pi})\right],&&\nonumber\\
 \phi_n=i(n+\frac{\alpha}{\pi})^{-\frac{1}{2}}e^{i\tau(n+\frac{\alpha}{\pi})}\sin \left[\sigma (n+\frac{\alpha}{\pi})\right].&&\end{eqnarray} 
We choose $0\leq\frac{\alpha}{\pi}<1$. Let $n_\alpha\equiv n+\frac{\alpha}{\pi}.$\\
The quantization of this free theory with nontrivial boundary conditions can be done using the method of~\cite{callan}.
Consider operators \be  {\cal O}_{11}=\frac{1}{2i}(i\stackrel{\leftrightarrow}{\partial}_\tau+\tan\alpha\:\delta(\pi-\sigma))\ee and \be  {\cal O}_{22}=\frac{1}{2}(i\stackrel{\leftrightarrow}{\partial}_\tau-\cot\alpha\:\delta(\pi-\sigma)). \ee  
It turns out that functions $\psi_n$ and $\phi_n $ are orthogonal in the following sense: \begin{eqnarray} \frac{1}{\pi}\int_0^\pi d\sigma\:\bar{\psi}_m{\cal O}_{11}\psi_n = \frac{1}{2}{\rm sign}\,n_\alpha\,\delta_{mn},&&\nonumber\\
 \frac{1}{\pi}\int_0^\pi d\sigma\:\bar{\phi}_m{\cal O}_{22}\phi_n = \frac{1}{2}{\rm sign}\,n_\alpha\,\delta_{mn}.&&\end{eqnarray}

Moreover, \begin{eqnarray}&
                         a^\dagger_m = \frac{\sqrt2}{\pi}{\rm sign}\,m_\alpha
			\int_0^\pi\ d\sigma \left\{ \left( \begin{array}{cc}
                         X_1, & X_2   
			\end{array} \right)
			\left(\begin{array}{cc}
			 {\cal O}_{11} & 0 	      \\
		         0	       & {\cal O}_{22}\\
			\end{array} \right)
		        \left(\begin{array}{c}
			 \psi_m \\ \phi_m
			\end{array} \right) \right\}, &\nonumber\\
		&	a_m = \frac{\sqrt2}{\pi}{\rm sign}\,m_\alpha
			\int_0^\pi\ d\sigma \left\{ \left( \begin{array}{cc}
                         \bar{\psi}_m, & \bar{\phi}_m 
			\end{array} \right)
			\left(\begin{array}{cc}
			 \bar{\cal O}_{11} & 0 	      \\
			 0	       & \bar{\cal O}_{22}\\
			\end{array} \right)
		        \left(\begin{array}{c}
			 X_1 \\ X_2
			\end{array} \right) \right\}. &\end{eqnarray}
Writing $a_n$ explicitly,
and substituting the expression for the momentum,\\ $P^\mu_\tau = \frac{1}{\pi}\partial_\tau X^\mu, $ we have: \begin{eqnarray}
a_m^\dagger =\displaystyle{\frac{1}{\sqrt2\pi}{\rm sign}\,m_\alpha\int_0^\pi d\sigma  (-m_\alpha }X_1(\sigma)\psi_m(\sigma)-i\pi P_1(\sigma)\psi_m(\sigma)+&&
\nonumber\\ 
im_\alpha X_2(\sigma)\phi_m(\sigma)-\pi P_2(\sigma)\phi_m(\sigma)).&&\label{eq:boscil}
\end{eqnarray}
Note that in this formula the boundary terms cancel.
We can now get the commutation relations. The result is: \begin{eqnarray}
&& [ a_m^\dagger , a_n^\dagger ]=0 \nonumber\\ 
&& [ a_m^\dagger , a_n ]={\rm sign}\,n_\alpha\,\delta_{mn}.
\end{eqnarray}
From these commutation relations we conclude that
\be  \begin{array}{l}
a_n=\left\{ \begin{array}{ll}
		\mbox{creation} & \mbox{if $n\geq0$} \\
		\mbox{destruction} & \mbox{if $n<0$}
		\end{array}
		\right. \\
a_n^\dagger=\left\{ \begin{array}{ll}
		\mbox{destruction} & \mbox{if $n\geq0$} \\
		\mbox{creation} & \mbox{if $n<0$}
		\end{array}
		\right.
  \end{array} \ee 
Now we find Virasoro operators, $L_m:$\\
\be  L_m=\frac{1}{\pi}\int_0^\pi(e^{im\sigma}:T_{++}:+e^{-im\sigma}:T_{--}:)d\sigma. \ee  Substituting the $T_{++}=\frac{1}{4}(\dot{X}+X^{'})^2,\;  T_{--}=\frac{1}{4}(\dot{X}-X^{'})^2$ 
we have \be  L_m=\frac{1}{2\pi}\int_0^\pi:(\dot{X}^2+X^{'2}):\cos m\sigma d\sigma +\frac{i}{\pi}\int_0^\pi:\dot{X}\cdot X^{'}:\sin m\sigma d\sigma. \ee
  The final expression is \be L_m=\sum_{n\in \cal Z}n_\alpha^\frac{1}{2}((n+m)_\alpha^\frac{1}{2})^*:a_n a^\dagger_{n+m}:.\ee Reality of $T_{\alpha\beta}$ requires that $L_m^\dagger=L_{-m}.$
To find the structure of the algebra of these operators, let's return to Poisson brackets, $[a,b]_{\rm pb}=-i[a,b].$ Also, we introduce operators $\alpha_n=\sqrt n_\alpha a_n.$ We then have \be L_n=\sum_{k\in \cal Z}\alpha^\dagger_{k+n}\alpha_k,\ee  \be  [\alpha_m^\dagger\alpha_n]_{\rm pb}=-i\; {\rm sign}\,n_\alpha\sqrt( n_\alpha^2)\,\delta_{mn}=-in_\alpha\delta_{mn}.\ee 
From this we find \be [L_m,L_n]_{\rm pb}=i(m-n)L_{m+n}.\ee 
Therefore the generators $L_n$ satisfy Virasoro algebra. Now let us return to the quantum mechanics. In general, the central term of this algebra is $A(n)=c_3n^3+c_1n.$ Consider \begin{eqnarray}
 &<0|[L_1,L_{-1}]|0>=&<0|\alpha_{-1}\alpha_0^\dagger\alpha^\dagger_{-1}\alpha_0|0>=\frac{\alpha}{\pi}\left(1-\frac{\alpha}{\pi}\right)=c_3+c_1, \nonumber\\
&<0|[L_2,L_{-2}]|0>=&<0|(\alpha^\dagger_0\alpha_{-2}+\alpha_1^\dagger\alpha_{-1})
(\alpha_{-2}^\dagger\alpha_0+\alpha_{-1}^\dagger\alpha_1)|0>=\nonumber
\\ &&=1-2\left(\frac{\alpha}{\pi}\right)^2+2\frac{\alpha}{\pi}=8c_3+2c_1. \end{eqnarray}
From this we find that
\be
A(n)=\frac{2}{12}n^3+\left(-\frac{2}{12}+\frac{\alpha}{\pi}-\left(\frac{\alpha}{\pi}\right)^2\right)n. \ee
The central charge can be brought to the standard form $\frac{D}{12}(n^3-n)$ by the redefinition \be L_0 \rightarrow L_0+\frac{\alpha}{2\pi}\left(1-\frac{\alpha}{\pi}\right).\ee Taking into account the transverse oscillators, an equation for a physical state is \be \left(L_0-1+\frac{\alpha}{2\pi}\left(1-\frac{\alpha}{\pi}\right)\right)|\psi>=0. \ee

Consider the superstring action in RNS formulation: \be  S=-\frac{1}{2\pi}\int_\Sigma d^2\sigma( \partial_\alpha X^\mu\partial^\alpha X_\mu-i\bar{\psi}^\mu\rho^\alpha\partial_\alpha\psi_\mu). \ee  Now we vary this action with supersymmetry transformations, \be  \delta\psi=-i\rho^\alpha\partial_\alpha X^\mu\varepsilon, \;\;\delta X^\mu=\bar{\varepsilon}\psi^\mu.\ee  Naturally, the bulk term of the variation vanishes, and we are left with a boundary term: \be \delta S=\frac{1}{2\pi}\int_{\partial\Sigma}d\sigma^\alpha\epsilon_{\alpha\beta}\bar{\varepsilon}\rho^\beta\rho^\gamma\partial_\gamma X^\mu\psi_\mu.\ee  Our goal is to determine fermionic boundary conditions if the bosonic ones are specified. To do this, we have to eliminate half of the $\partial_\gamma X^\mu$ at each boundary, using the boundary conditions for bosons. Specifically, in our case, \be \delta S=\frac{1}{2\pi}\int_{\sigma =0,\pi} d\tau\bar{\varepsilon}\rho^1\rho^\gamma\partial_\gamma X^\mu\psi_\mu.\ee  At 
$\sigma =0,$ \be \partial_1 X^1=0,\;\;\partial_0X^2=0.\ee  Writing the contribution for the first two coordinates only, \be \delta S=\int_{\sigma=0}d\tau( \bar{\varepsilon}\rho^1\rho^0\psi_1\partial_0 X^1+\bar{\varepsilon}\rho^1\rho^1\psi_2\partial_1 X^2)=0.\ee  Since $\partial_0 X^1$ and $\partial_1 X^2$ are arbitrary, we conclude that at this boundary \begin{eqnarray}
&\bar{\varepsilon}\rho^1\rho^0\psi_1=0 \nonumber \\
&\bar{\varepsilon}\rho^1\rho^1\psi_2=0.\end{eqnarray}  These are two equation for  four variables: $
\varepsilon^{-}\psi^{+}_{1,2},\; \varepsilon^{+}\psi^{-}_{1,2}$. To complete the system we have to use the boundary conditions on the other end, $\sigma=\pi$: \begin{eqnarray}
& \psi_1\partial_\sigma X^1+\psi_2\partial_\sigma X^2=(-\tan\alpha\psi_1+\psi_2)\partial_\sigma X^2\nonumber \\
&\psi_1\partial_\tau X^1+\psi_2\partial_\tau X^2=(\psi_1+\tan\alpha\psi_2)\partial_\tau X^1,\end{eqnarray}  where $\partial_\sigma X^2$ and $\partial_\tau X^1$ are now arbitrary. This gives the last two equations: \begin{eqnarray} 
&\bar{\varepsilon}\rho^1\rho^0( \psi_1+\tan\alpha\psi_2)=0 \nonumber\\ \nopagebreak
& \bar{\varepsilon}\rho^1\rho^1( \psi_2-\tan\alpha\psi_1)=0. \end{eqnarray}
Using the fact that $\varepsilon$ is the same at both ends and denoting $-i\varepsilon^{-}\equiv\gamma$ and $i\varepsilon^{+}\equiv\beta$, we have the following equations:
\begin{eqnarray}
&  \beta\psi_1^{+}(\tau+\pi)-\gamma\psi_1^{-}(\tau-\pi)=-\tan\alpha(\beta\psi_2^{+}(\tau+\pi)-\gamma\psi_2^{-}(\tau-\pi)) \nonumber \\ 
& (\beta\psi_1^{+}(\tau+\pi)+\gamma\psi_1^{-}(\tau-\pi))\tan\alpha=\beta\psi_2^{+}(\tau+\pi)+\gamma\psi_2^{-}(\tau-\pi)\nonumber \\
&\gamma\psi_2^{-}(\tau)+\beta\psi_2^{+}(\tau)=0\nonumber \\ 
&\gamma\psi_1^{-}(\tau)-\beta\psi_1^{+}(\tau)=0.\end{eqnarray} 
Now we observe that they exactly coincide with equations~(\ref{eq:bbound}) for bosonic coordinates provided if we identify \begin{eqnarray}
& \dot{X}_\mu^{R}(\tau-\sigma)\Leftrightarrow\gamma\psi_\mu^{-}(\tau-\sigma) \nonumber \\
& \dot{X}_\mu^{L}(\tau+\sigma)\Leftrightarrow\beta\psi_\mu^{+}(\tau+\sigma). \end{eqnarray}  The choice R versus NS boundary conditions
translates into the ambiguity of the choice of the order of product $\gamma\psi^{-}=-\psi^{-}\gamma$ versus $\beta\psi^{+}.$ The further computations are the same as for bosons, in particular, $\psi_2^{-}$ satisfies the equation 
\be \psi_2^{-}(\tau+4\pi)\mp 2\cos 2\alpha\psi_2^{-}(\tau+2\pi)+\psi_2^{-}(\tau)=0, \ee 
where the upper sign corresponds to R and lower to NS boundary conditions.
The solution for such an equation is again the same as for $\dot{X}^2_R$ equation, except that $c(\tau+2\pi)=-c(\tau)$ in NS sector. Now we can write the mode expansion: \begin{eqnarray}
& \psi_1(\tau,\sigma)=\frac{1}{2}\sum_{n\in \cal A}\left[-id_n 
\left(\begin{array}{c}
e^{i n_\alpha(\tau-\sigma)} \\
\pm e^{i n_\alpha(\tau+\sigma)}
\end{array} \right)+cc
\right]\nonumber \\ 
& \psi_2(\tau,\sigma)=\frac{1}{2}\sum_{n\in \cal A}\left[d_n 
\left(\begin{array}{c}
e^{i n_\alpha(\tau-\sigma)} \\
\mp e^{i n_\alpha(\tau+\sigma)}
\end{array} \right)
+cc\right],\end{eqnarray}  where the upper sign and $\cal A=\cal Z$ corresponds to R and lower sign and $\cal A=\cal Z+\em\frac{1}{2}$ corresponds to NS sector.
In the Ramond sector consider the expressions \begin{eqnarray} 
&\psi^{+}_1-\psi^{-}_1=\sum_{n \in \cal Z}\left[-in_\alpha^{\frac{1}{2}}d_n\phi_n+cc\right]\nonumber \\ 
&\psi^{-}_2-\psi^{+}_2=\sum_{n \in \cal Z}\left[n_\alpha^{\frac{1}{2}}d_n\psi_n+cc\right] \end{eqnarray} 
Therefore, we can find $d_n$ using formula~(\ref{eq:boscil}) for bosonic coordinates, replacing $X_1\rightarrow\psi^{-}_2-\psi^{+}_2,$ $X_2\rightarrow\psi^{+}_1-\psi^{-}_1.$ Similarly to bosonic case, we find commutation relations.
The result is \be 
\{d_m , d_n\}=0,\;\;\{d_m^\dagger , d_n\}=\delta_{mn}.\ee 
From here we can choose
\be  \begin{array}{l}
d_n=\left\{ \begin{array}{ll}
		\mbox{creation} & \mbox{if $n\geq0$} \\
		\mbox{destruction} & \mbox{if $n<0$}
		\end{array}
		\right. \\
d_n^\dagger=\left\{ \begin{array}{ll}
		\mbox{destruction} & \mbox{if $n\geq0$} \\
		\mbox{creation} & \mbox{if $n<0$}
		\end{array}
		\right.
\end{array} \ee 
This identification correctly reproduces usual creation and annihilation operators in $\alpha=0$ limit:
\begin{eqnarray}
&d_n^\dagger=\frac{1}{\sqrt{2}}(d_n^2-id_n^1), \nonumber \\
&d_n=\frac{1}{\sqrt{2}}(d_{-n}^2+id_{-n}^1). 
\end{eqnarray}
The fermionic part of Virasoro operators is
\be L_m^{(d)}=\frac{1}{2}\sum_{n \in \cal Z}(m+2n_\alpha):d_nd^\dagger_{m+n}:. \ee
Here, unlike NN or DD strings, the worldsheet hamiltonian $L_0$ contains $d_0$ and $d^\dagger_0$, and therefore,
\be 
\{L_0,d_0^\dagger\}=\frac{\alpha}{\pi}d_0^\dagger. \ee
In particular,
\be
L_0d^\dagger_0|0>=\frac{\alpha}{\pi}d_0^\dagger|0>, \ee
meaning that part of the ground state degeneracy is now removed.
Similar computation to the bosonic case shows, that the Poisson brackets of Virasoro operators obey classical Virasoro algebra. Now we can find the central charge.\be
\begin{array}{l}
<0|[L_1^{(d)},L_{-1}^{(d)}]|0>=\frac{1}{4}\left(2\frac{\alpha}{\pi}-1\right)^2<0|d_{-1}d_0^\dagger d_0\alpha_{-}^\dagger|0>=\frac{1}{4}-\frac{\alpha}{\pi}+\left(\frac{\alpha}{\pi}\right)^2, \\
<0|[L_2^{(d)},L_{-2}^{(d)}]|0>=\\
=<0|((\frac{\alpha}{\pi}-1)d_{-2}d_0^\dagger+\frac{\alpha}{\pi}d_{-1}d_1^\dagger)
((\frac{\alpha}{\pi}-1)d_0d_{-2}^\dagger+\frac{\alpha}{\pi}d_1d_{-1}^\dagger)|0>=\\
=1+2\left(\frac{\alpha}{\pi}\right)^2-2\frac{\alpha}{\pi}. \end{array}\ee
Adding the bosonic contribution, we get
\be \left\{
\begin{array}{l}
c_3+c_1=\frac{1}{4}\nonumber\\
8c_3+2c_1=2 \end{array}
\right. \ee
Therefore, the central charge is unaffected, $A(n)=\frac{D}{8}n^3=\frac{2}{8}n^3.$ Let us now turn to NS sector. Examining the equations shows that the commutation relations and Virasoro generators are the same as in R sector, with sums over integers replaced by the sums over half-integers. This changes the central
 charge. 
\be
\begin{array}{l}
<0|[L_1^{(b)},L_{-1}^{(b)}]|0>=\left(\frac{\alpha}{\pi}\right)^2, \\
<0|[L_2^{(d)},L_{-2}^{(d)}]|0>=\frac{1}{2}+2\left(\frac{\alpha}{\pi}\right)^2. \end{array}\ee
The central charge is \be
A(n)=\frac{1}{4}n^3+\left(-\frac{1}{4}+\frac{\alpha}{\pi}\right)n. \ee
Again, this is brought into the standard form $\frac{D}{8}(n^3-n)$ by the redefinition \be L_0 \rightarrow L_0+\frac{\alpha}{2\pi}.\ee
This means that after taking into account the contribution of 8 transverse modes, the normal ordering constant in NS sector becomes
\be a=\frac{1}{2}-\frac{\alpha}{2\pi}.\ee
We are looking for the quantity 
\be
{\cal A}={\rm Tr}\log\frac{1}{L_0}
\ee
where $L_O$ includes the normal ordering constant. In bosonic case 
\be
{\cal A}=-\Gamma (0)+\int_0^\infty\,\frac{dt}{t}{\rm Tr}e^{-tL_0}.
\label{eq:btr}
\ee
Using the bosonic part of the expression given in~\cite{pol1} and substituting $X^1,\,X^2$ contribution and new normal ordering constant, we get
\be
\begin{array}{c}
{\cal A}={\rm T}\int_0^\infty\,\frac{dt}{t}(2\pi t)^{-\frac{1}{2}}e^{-t(\frac{Y^2}{2\pi^2}-1+\frac{\alpha}{2\pi}(1-\frac{\alpha}{\pi}))}
(1-q^{2\frac{\alpha}{\pi}})^{-1}\times\\ \\ \times
\prod_{n=1}^{\infty}(1-q^{2n})^{-22}(1-q^{2(n-\frac{\alpha}{\pi})})^{-1}
(1-q^{2(n+\frac{\alpha}{\pi})})^{-1}.
\end{array}
\ee
In this formula, $T$ is the total time of interaction, or the volume of time-like NN direction, $Y$ is the minimal separation between the branes and $q=e^{-\frac{t}{2}}$.

In the superstring case, ${\rm Tr}\,1=0$ because of the supersymmetry and ${\cal A}$ is finite, which after taking into account the GSO projection and the sum over spin structures takes the form 
\be
{\cal A}=\int_0^\infty\,\frac{dt}{t}\sum_{a,b=0,1}c_{ab}{\rm Tr}(-)^{bF}e^{-tL_0^{(a)}}.
\ee
Here $F$ is the world-sheet fermion number $a=0,1$ labels NS and R sectors respectively and and the coefficients in the sum over spin
structures are $c_{00}=-c_{01}=-c_{10}=\pm c_{11}=\frac{1}{2}.$
We obtain
\be
\begin{array}{c}
{\cal A}={\rm T}\int_0^\infty\,\frac{dt}{t}(2\pi t)^{-\frac{1}{2}}e^{-\frac{tY^2}{2\pi^2}}
(1-q^{2\frac{\alpha}{\pi}})^{-1}\times \\ \\ \times
\prod_{n=1}^{\infty}(1-q^{2n})^{-6}(1-q^{2(n-\frac{\alpha}{\pi})})^{-1}
(1-q^{2(n+\frac{\alpha}{\pi})})^{-1}\times \\ \\ \times
\frac{1}{2}\left\{-8(1+q^{2\frac{\alpha}{\pi}})\prod_{n=1}^\infty(1+q^{2n})^6(1+q^{2(n+\frac{\alpha}{\pi})})
(1+q^{2(n-\frac{\alpha}{\pi})})+ \right. \\ \\+
q^{-1+\frac{\alpha}{\pi}}\prod_{n=1}^\infty(1+q^{2n-1})^6(1+q^{2(n+\frac{\alpha}{\pi})-1})
(1+q^{2(n-\frac{\alpha}{\pi})-1})+  \\ \\ \left. +
q^{-1+\frac{\alpha}{\pi}}\prod_{n=1}^\infty(1-q^{2n-1})^6(1-q^{2(n+\frac{\alpha}{\pi})-1})
(1-q^{2(n-\frac{\alpha}{\pi})-1}) \right\}.
\end{array}
\ee
In this formula the first term in curly brackets stands for R sector while the second and the third - for NS without and with $(-)^F$ respectively. We can rewrite the amplitude using $\theta$-functions defined in the appendix 8.A of~\cite{GSW}. The result is
\be
\begin{array}{c}
{\cal A}=2i{\rm T}\int_0^\infty\,\frac{dt}{t}(2\pi t)^{-\frac{1}{2}}
e^{-t\frac{Y^2}{2\pi^2}}\left(\frac{2\pi}{\theta_1'(0|\tau)}
\right)^3\theta_1^{-1}(\nu|\tau)
\times \\ \\ \times \left\{
(-\theta_2^3(0|\tau)\theta_2(\nu|\tau)+\theta_3^3(0|\tau)\theta_3(\nu|\tau)-\theta_4^3(0|\tau)\theta_4(\nu|\tau) 
\right\} . \label{eq:ampl}
\end{array}
\ee
Here $\tau=\frac{it}{2\pi}$ and $\nu=-\frac{it}{2\pi}\frac{\alpha}{\pi}.$

A similar result was given in the paper~\cite{iran}. However, it turns out to be essential that the normal ordering constant in NS sector is angle dependent to get the structure in curly brackets of~(\ref{eq:ampl}) so that the amplitude would vanish by the ``abstruse identity'' for parallel branes. This amplitude is infinite for $\alpha=\pi$ due to the contribution of $\theta_1(-\tau | \tau )$. Explicitly,
\be
{\cal A}=2i{\rm T}\int_0^\infty\,\frac{dt}{t}(2\pi t)^{-\frac{1}{2}}\frac{e^{-t\frac{Y^2}{2\pi^2}}}{\sin (\pi\nu)}f(\tau,\nu),
\ee
where the divergent part is displayed.
It would also be infinite for parallel branes, but the expansion of the fermionic contribution in~(\ref{eq:ampl}) starts as $\alpha^4$ for small alpha. These infinities are due to the fact that the amplitude describes the interaction between the entire D-strings, which is reflected by the fact that in the usual cases of parallel D-branes it is multiplied by the infinite volume $V_D$. A finite quantity is the potential per unit length, ${\cal V},$ which can be defined by
\be
{\cal A}=\frac{1}{\sqrt{2\pi^2}}\int_{-\infty}^{\infty}\, dl\; {\cal V}(\alpha,r^2=Y^2+\sin^2\alpha\, l^2),
\ee
where $r$ denotes the minimal distance between the line element $dl$ of the tilted D-string (at $\sigma=\pi$) and the D-string at $\sigma=0$ and the integration is along the tilted D-string. This potential is the analog of the effective 
potential for moving D0-branes~\cite{douglas}.
From this definition, 
\be
{\cal V}(\alpha,r^2)=2i{\rm T}\int_0^\infty\,\frac{dt}{t}(2\pi )^{-\frac{1}{2}}e^{-t\frac{r^2}{2\pi^2}}\frac{\sin\alpha}{\sin (\pi\nu)}f(\tau,\nu).
\ee
As a check we can find a force per unit length at large $Y$ for antiparallel D-strings:
\be
{\cal F}=\frac{d{\cal V}}{dY}\sim\frac{1}{Y^7}.
\ee
In derivation of this formula one should notice that at $\alpha=\pi$ only $-2\theta_4^4(0|\tau)$ remains in the fermionic part of the potential after using Jacobi identity and then small $t$ expansion can be performed after using modular transformation properties when $\tau\rightarrow -\frac{1}{\tau}.$
\subsection*{Acknowledgments}

I would like to thank Samir Mathur for his guidance through this problem,
as well as Constantin Bachas and Slava Zhukov for useful discussions. This work is supported in part by funds provided by the U.S. Department of Energy (D.O.E.) under cooperative research agreement DE-FC02-94ER40818.

\end{document}